\documentclass[paper,11pt]{JHEP}
\usepackage[centertags]{amsmath}
\usepackage{amsfonts} \usepackage{amssymb} \usepackage{amsthm}
\usepackage{graphicx}
\usepackage{caption}
\newcommand{\Rw}{R_W}
\newcommand{\Rp}{R_P}

\def\one{{\rm 1\kern -.9mm l}}                             %

\def\beq{\begin{equation}}
\def\eeq{\end{equation}}
\def\beq{\begin{equation}}
\def\eeq{\end{equation}}
\def\beqa{\begin{eqnarray}}
\def\eeqa{\end{eqnarray}}
\newcommand{\eqa}{\begin{eqnarray}}
\newcommand{\ena}{\end{eqnarray}}
\newcommand{\eq}[1]{eq. (\ref{#1})}

\def\ii{\mathrm{i}}
\def\ee{\mathrm{e}}
\newcommand{\Z}{\mathbb{Z}}

\newcommand{\cL}{\mathcal{L}}

\title{The Lorentz-invariant boundary action of the confining string and its universal contribution\\
to the inter-quark potential}
\author{M. Bill\'o, M. Caselle, F. Gliozzi, M. Meineri, R. Pellegrini
\\
\vskip 0.2cm
Dipartimento di Fisica, Universit\`a di Torino\\
and Istituto Nazionale di Fisica Nucleare - sezione di Torino \\
Via P. Giuria 1, I-10125 Torino, Italy\\
\vspace{0.25cm}
\email{billo,caselle,gliozzi,pellegri@to.infn.it} 
}
\abstract{We study the boundary contribution to the low energy effective action of the open string describing the confining flux tube in gauge theories. The form of the boundary terms is strongly constrained by the requirement of Lorentz symmetry, which is spontaneously broken by the formation of a long confining flux tube in the vacuum. Writing the boundary action as an expansion in the 
derivatives of the Nambu-Goldstone modes describing the transverse fluctuations of the string, we single out and  put in a closed form the first 
few Lorentz invariant boundary terms. We also evaluate the leading deviation 
from the Nambu-Goto string produced by the boundary action on the vacuum expectation value of the Wilson loop and we test this prediction  in the 
3d Ising gauge model. Our simulation attains 
a level of precision which is sufficient to test the  contribution of 
this term.
}
\preprint{DFTT/02/2012}
\begin{document}

\section{Introduction}
\label{sec:intro}
The confinement of quarks is one of the most fundamental features of the strong interactions 
and our understanding of it is far from being complete.
However, even if the origin of this non-perturbative phenomenon remains unclear, it is still possible to quantitatively understand some of its consequences. 
In particular the vacuum state in presence of a pair of quark sources is 
believed to be dominated, for large quark separations, by the formation of a thin flux tube which generates the linear rising of the confining potential. 
Clearly this configuration breaks spontaneously  some of 
the space-time symmetries of the underlying gauge theory. 
The resulting massless Nambu-Goldstone bosons are the natural degrees of freedom to be taken into account in the low-energy regime of 
the theory. They define the embedding of the string-like flux tube  in the space-time. As we shall see, one may obtain important  consequences of spontaneous  breaking of the space-time symmetries, even if the dynamics leading to the formation of the confining flux tube is unknown.  

As first proposed by L\"uscher, Symanzik and Weisz \cite{Luscher:1980fr}, 
the effective action of  the color flux tube connecting a static quark-antiquark pair in a $d$-dimensional confining Yang-Mills theory is a two-dimensional action describing the world-sheet swept out by the flux string in its time evolution. The leading term in the infrared limit is a free-field action for the $d-2$  Nambu-Goldstone modes resulting from the spontaneous breaking of  the translation invariance in the transverse directions. The ensuing universal string fluctuation effects, namely, the leading correction to the linear term in the static potential \cite{Luscher:1980ac} and the  growth of the flux tube 
cross-section with the logarithm of the interquark distance 
\cite{Luscher:1980iy} were first unambiguously observed many years ago in 
the $\Z_2$ gauge theory in three dimensions  
\cite{{Hasenbusch:1992zz},{Caselle:1995fh},{Caselle:1996ii}} and more recently in non-abelian Yang-Mills theories 
\cite{{Lucini:2001nv},{Luscher:2002qv},{Caselle:2004er},{Gliozzi:2010zv},{Gliozzi:2010zt}}. 

In recent years, the great numerical accuracy of Monte Carlo simulations 
allowed to check and analyze also the subleading corrections of the free-field 
action. Actually the Lagrangian of the low-energy effective theory consists 
of all terms respecting the internal and space-time symmetries of the system.
The action of the effective string theory can be written as low energy expansion in the number of derivatives of the Nambu-Goldstone fields. In particular the 
first few terms of the effective action $S$ for an open string stretched between fixed ends, for instance Polyakov lines, is \cite{Luscher:2004ib}
 \beq
S=S_{cl}+\frac\sigma2\int d^2\xi\left[\partial_\alpha X\cdot\partial^\alpha X+
c_2(\partial_\alpha X \cdot\partial^\alpha X)^2
+c_3(\partial_\alpha X \cdot\partial_\beta X)^2+\dots\right]+S_b\,,
\label{action}
\eeq
where the classical term $S_{cl}$  describes the usual perimeter-area term.
The Nambu-Goldstone fields $X_i(\xi_0,\xi_1)$  $(i=1,\dots,d-2)$ 
parametrize the displacements orthogonal to the surface of minimal area representing the configuration around which we expand and
$\xi_0,\xi_1$ are the world-sheet coordinates. $S_b$ is the boundary contribution characterizing the open string. If the boundary is a Polyakov line in the $\xi_0$ 
direction placed at $\xi_1=0$, on which we assume  Dirichlet boundary conditions
$X_i(\xi_0,0)=0$, the first few terms are
\beq
S_b=\int d\xi_0 \left[b_1\partial_1 X\cdot \partial_1 X+b_2\partial_1\partial_0 X\cdot \partial_1\partial_0 X+b_3(\partial_1 X\cdot \partial_1 X)^2+
\dots\right]\,.
\label{bounda}
\eeq
As first observed in 2004 by L\"uscher and Weisz \cite{Luscher:2004ib} the coefficients $c_i$ and $b_i$ of the above expansion should satisfy some consistency constraints; they were obtained by the comparison of the string partition function in different channels (``open-closed string duality''). These results were further generalized by Ref. \cite{Aharony:2009gg}. It was also realized 
\cite{{Meyer:2006qx},{aks},{Aharony:2010cx}} that the crucial ingredient of these 
constraints is the Lorentz symmetry of the underlying Yang-Mills theory.
Indeed, even if the complete $SO(1,d-1)$  
invariance is broken by the classical configuration around which we 
expand, the effective action should still respect this symmetry through a 
non-linear realisation in terms of transverse fields $X_i$. In this way 
it was shown \cite{{aks},{Aharony:2010cx},{Gliozzi:2011hj}} that the terms with
 only first derivatives  coincide with the Nambu-Goto action to all orders in the derivative expansion. The first allowed correction to the Nambu-Goto 
action turns out to be the the six derivative term  \cite{Aharony:2009gg}
\beq
c_4\left(\partial_\alpha\partial_\beta X\cdot\partial^\alpha\partial^\beta X
\right)\left(\partial_\gamma X\cdot\partial^\gamma X\right)
\eeq
with arbitrary coefficient $c_4$; however this term is non-trivial only 
when $d>3$. For $d=3$ the first non-trivial deviation of the Nambu-Goto action 
is an eight-derivative term and it has been recently shown \cite{Aharony:2011gb}, using the recursion relations generated by the non-linear Lorentz 
transformations, that it generates a geometric term proportional to the 
squared curvature of the induced metric on the world-sheet. 
The fact that the first deviations from the Nambu-Goto string are of 
high order, especially in $d=3$, explains why in early Monte Carlo 
calculations \cite{{Caselle:1994df},{Caselle:2005xy},{Caselle:2006dv}}  a good agreement with the Nambu-Goto string was observed. The leading deviations of the effective string spectrum were explicitly calculated in 
\cite{Aharony:2010db} and they turn out to be consistent with lattice simulations for long closed confining strings \cite{{Athenodorou:2011rx},{Athenodorou:2010cs}}.

In this paper we deal with open effective strings both from the point of 
view of theoretical investigation and numerical calculations. 
In \cite{Aharony:2010cx} it was already observed that in the boundary action 
(\ref{bounda}) only the $b_2$ term passes the first test of Lorentz invariance 
(the vanishing of $b_1$ was previously proven in \cite{Luscher:2004ib}). Here
we complete the proof by solving the recursion relations dictated by the requirement of Lorentz invariance. The resulting expansion can be written in a closed form, yielding
\beq
b_2\int d\xi_0 \left[
\frac{\partial_0\partial_1 X\cdot\partial_0\partial_1 X}{1+\partial_1 X\cdot\partial_1X}-
\frac{\left(\partial_0\partial_1 X\cdot\partial_1 X\right)^2}
{\left(1+\partial_1 X\cdot\partial_1X\right)^2}\right]\,.
\label{firstb}
\eeq 
Using the same technique we worked out also the next few boundary invariants. 
For instance, the complete form of that generated by the six derivative term 
$\partial_0^2\partial_1 X\cdot\partial_0^2\partial_1X$  in 
$d=3$ is $b_4^{(1)}\int d\xi_0\cL_{b,4}$, where
\beq
\cL_{b,4}=\frac{\left(\partial^2_0\partial_1 X\right)^2}{\left(1+
(\partial_1 X)^2\right)^2}-4\frac{\partial_1X\partial^2_0\partial_1 X
\left(\partial_0\partial_1X\right)^2}{\left(1+
(\partial_1 X)^2\right)^3}+4\frac{\left(\partial_0\partial_1X\right)^4
\left(\partial_1 X\right)^2}{\left(1+
(\partial_1 X)^2\right)^4}~,
\label{six}
\eeq
with an arbitrary coefficient $b_4^{(1)}$ (this nomenclature will become clear later). 

The lowest order term in (\ref{firstb}) represents the first deviation from the Nambu-Goto string in $d=3$, 
it is thus important to look for the contribution of this term to physical observables in lattice gauge theories. 
A first test of the effect of this term on  the open string energy levels has been already performed in 
\cite{Brandt:2010bw}, in the case of $\mathrm{SU}(2)$ gauge theory. In the present paper we probe the  contribution of this boundary term to the vacuum expectation value of Polyakov loop correlators and of Wilson
loops, using the three-dimensional $\Z_2$ gauge model. We shall see that this contribution is of the same order of magnitude of the subleading effective string corrections to the interquark potential
 which were recently studied both using Polyakov loop correlators \cite{{Caselle:2005xy},{Caselle:2010pf}} and Wilson loops \cite{Billo:2011fd} 
 and thus cannot be neglected when performing high precision tests of the effective string.
We use the data obtained from our simulation of Polyakov loop correlators  to fit the coefficient $b_2$ of the boundary correction. We then use this value to predict the Wilson loop behaviour;  we have no free parameter left in this comparison, so the 
remarkable agreement that we find with the data  represents a strong consistency check of our treatment of the boundary correction given in \eq{firstb}. 

 As it is easy to guess, these boundary corrections are more important in the Wilson case than in the Polyakov loop correlators case. 
 In particular, in the Wilson loop case they perfectly explain the deviations recently observed in \cite{Billo:2011fd} 
 between effective string predictions and numerical results.
 It is possible that they could also explain the subtle, higher order, deviations recently observed using Polyakov loop correlators \cite{Caselle:2010pf}, 
 however such a test, which would require a much stronger numerical effort is beyond the scope of the present paper.

\section{ The boundary action}
 The confining string action  can be thought of as the
effective low energy action which is obtained integrating over all the massive 
degrees of freedom of  Yang-Mills theory in presence of a pair of static sources. As already emphasized in the Introduction, the formation of a confining flux tube spontaneously 
breaks the transverse translational as well as the Lorentz (or rotational) invariance of the bulk space-time. This symmetry is not manifest in the static gauge, where one uses only the physical degrees of freedom, associated to the transverse
fluctuations $X_i(\xi_0,\xi_1)$ $(i=1,\dots d-2)$.  However the effective string action should still respect this symmetry through a 
non-linear realisation.

\begin{figure}[htm]
\begin{center}
\begingroup%
  \makeatletter%
  \providecommand\color[2][]{%
    \errmessage{(Inkscape) Color is used for the text in Inkscape, but the package 'color.sty' is not loaded}%
    \renewcommand\color[2][]{}%
  }%
  \providecommand\transparent[1]{%
    \errmessage{(Inkscape) Transparency is used (non-zero) for the text in Inkscape, but the package 'transparent.sty' is not loaded}%
    \renewcommand\transparent[1]{}%
  }%
  \providecommand\rotatebox[2]{#2}%
  \ifx\svgwidth\undefined%
    \setlength{\unitlength}{220bp}%
    \ifx\svgscale\undefined%
      \relax%
    \else%
      \setlength{\unitlength}{\unitlength * \real{\svgscale}}%
    \fi%
  \else%
    \setlength{\unitlength}{\svgwidth}%
  \fi%
  \global\let\svgwidth\undefined%
  \global\let\svgscale\undefined%
  \makeatother%
  \begin{picture}(1,0.44134771)%
    \put(0,0){\includegraphics[width=\unitlength]{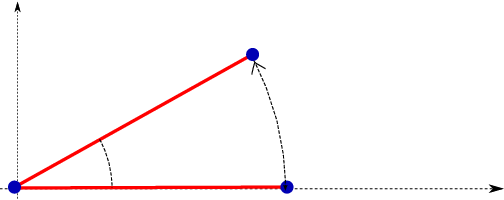}}%
\put(0.04756143,0.35862847){\makebox(0,0)[lb]{\smash{$X^i$}}}%
\put(0.5581544,-0.03){\makebox(0,0)[lb]{\smash{$P(R)$}}}%
 
\put(0.9,-0.03){\makebox(0,0)[lb]{\smash{$\xi^1$}}}%
  \put(0.0478757,-0.03){\makebox(0,0)[lb]{\smash{
$P^*(0)$}}}%
   
\put(0.23,0.07){\makebox(0,0)[lb]{\smash{$\epsilon$}}}

\put(0.28,0.03100007){\makebox(0,0)[lb]{\smash{string}}}%
  \end{picture}%
\endgroup%
\end{center}
\caption{Rotation of the Polyakov correlator around the $\xi_0$ axis 
.}
\label{Figure:1}
\end{figure}
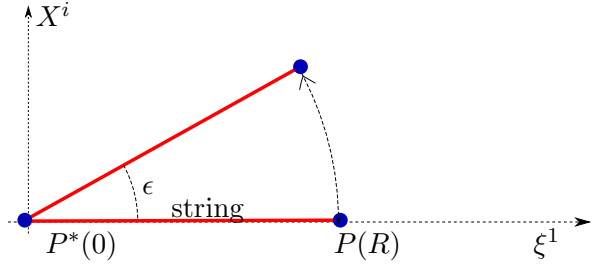

To make  the discussion concrete and explicit, we specialize now to the case 
where the string action describes the vacuum expectation value of the 
correlator of a pair of Polyakov loops $\langle P(0)^*P(R)\rangle$ parallel to the 
$\xi_0$ axis and located respectively at $\xi_1=0$ and $\xi_1=R$. The string stretched between these two Polyakov loops fluctuates in the $d-2$ transverse directions, orthogonal to the plane $[\xi_0,\xi_1]$. 
Rotate now this system of an infinitesimal angle $\epsilon$ 
around the $\xi_0$ axis in the plane $[\xi_1, i]$  as shown in 
Figure \ref{Figure:1}. Of course this transformation in a Lorentz invariant theory keeps the physics of the system invariant;  however, it modifies the world-sheet coordinates $\xi_0,\xi_1$.The linear transformation of the transverse fields $X_j$ generated by the rotation should therefore be followed by a reparametrization 
which puts again the system in the static gauge. As a result the string action should be invariant under the infinitesimal transformation
\beq
\delta\,X^j=
-\epsilon\,\delta^{i\,j} \xi_1-
\epsilon\, X^i\partial_1 X^j~,~~~(i,j=1,\dots d-2).
\label{rota}    
\eeq   
When it is applied to a term with $m$ derivatives and 
$n$ transverse fields $X_j$, schematically $\partial^m X^n$, it generates terms with the same value of the difference $m-n$, called ``scaling'' of the given term
\cite{Aharony:2011gb}. When we apply the transformation (\ref{rota}) to the 
boundary term $\partial_0\partial_1X\cdot\partial_0\partial_1X $ of 
(\ref{bounda}), which has scaling 2, it generates a tower of all possible 
terms of the same scaling value. Thus the Lagrangian density $\cL_{b,2}$ of 
all boundary terms of scaling 2 contributing to the boundary action 
$S_{b,2}=b_2\,\int d\xi_0\cL_{b,2}$ 
\beq
\cL_{b,2}=\sum_{k=0}^\infty\left[\alpha_k
\partial_1\partial_0X\cdot\partial_1\partial_0X(\partial_1X\cdot\partial_1X)^k
+\beta_{k+1}(\partial_1\partial_0X\cdot\partial_1X)^2(\partial_1X\cdot
\partial_1X)^k \right].
\label{scaling2}
\eeq
In listing these terms we eliminated, as is usual in this context 
\cite{Luscher:2004ib,Aharony:2009gg,Aharony:2010cx,Aharony:2011gb},
terms proportional to the equation of motion or its derivatives, since
these terms can be reabsorbed by a field redefinition; it is sufficient
to use the equation of motion of the free field theory, in view of the fact
that corrections to this will generate terms of higher order that can be
ignored since we list the most general terms at each order anyway. 

The invariance of $\cL_{b,2}$ under (\ref{rota}) dictates the following
recursion relations among the $\alpha_k$ and $\beta_k$ coefficients:
\beq
\alpha_k+\alpha_{k+1}=0~~,~~(k+1)\,\beta_k+k\,\beta_{k+1}=0~~,~~
\alpha_k+\beta_k+\beta_{k+1}=0\,.
\eeq 
The solution of these recursion relations depends on a single free parameter
and yields at once Eq.(\ref{firstb}), which for $d=3$ becomes simply
\beq
\cL_{b,2} \stackrel{d=3}{=} \frac{\left(\partial_0\partial_1X\right)^2}
{\left(1+(\partial_1X)^2\right)^2}\,.
\label{firstb3}
\eeq
It is easy to verify that \eq{firstb}, hence in particular \eq{firstb3}, is not a total derivative, because this is possible only if each term in (2.2) is a total derivative, but direct
inspection shows  that this is not the case. An indirect proof of this fact is that the vacuum expectation value of the first term of (2.2), calculated in
section 3, is different from zero. 

It is not difficult to enlarge this game to terms of higher scaling.
There is no invariant term of scaling 3; in fact, it is possible to prove that there are no invariant boundary terms of odd scaling. There are two invariant terms of scaling 4. One is the invariant  $\cL_{b,4}$ whose form in $d=3$ was anticipated in (\ref{six}); we sketch its derivation in appendix \ref{appb}. The other is simply  $(\cL_{b,2})^2$, the square of the scaling 2 invariant 
(\ref{firstb3}). There are five invariants of scaling 6, but two of them are simply  the product of Lorentz invariants of lower degree.  
In conclusion, the most general boundary action, up to terms of scaling 8, 
can be written in the form
\beq
S_b=\int d\xi_0\left[b_2\cL_{b,2}+ b_4^{(1)}\cL_{b,4}+b_4^{(2)}\,
(\cL_{b,2})^2+\sum_{i=1}^3 b_6^{(i)}\cL_{b,6}^i
+b_6^{(4)}(\cL_{b,2})^3+b_6^{(5)}\cL_{b,2}\cL_{b,4}\right],
\eeq
with $b_k^{(i)}$  arbitrary coefficients. The general expressions of 
$\cL_{b,k}^i$ for arbitrary space-time dimensions $d$  are reported in 
appendix \ref{appb}.

\section{Effects of the first boundary correction in specific geometries}
\label{sec:bcwilson}
In this section we investigate the effect on Polyakov loop correlators and Wilson 
loop expectation values of the first term in the boundary component of the string effective action \eq{firstb}.
We aim for comparison with the lattice simulations discussed in next Section, hence we focus on the
$d=3$ case, where there is a single transverse embedding field
and the action reduces, as noted in \eq{firstb3}, to
\begin{equation}
 \label{ac3}
 S_{b,2} = b_2 \int_{\partial \Sigma} \frac{(\partial_0\partial_1 X)^2}{\left(1 + (\partial_1 X)^2\right)^2}
 = b_2 \int_{\partial \Sigma}\left\{(\partial_0\partial_1 X)^2 
 - 2  (\partial_0\partial_1 X)^2 (\partial_1 X)^2 + \ldots~.
 \right\}
\end{equation}

In our simulations we will be able to check the effect of the leading term in the  
derivative expansion of this boundary action, namely 
\begin{equation}
\label{derexpsb1}
S_{b,2}^{(1)} = b_2 \int_{\partial \Sigma} (\partial_0\partial_1 X)^2~
\end{equation}
and to fit reliably the value of $b_2$.
The first subleading correction in \eq{ac3} appears at the same order in the derivative expansion
as the leading term arising from the possible scaling-weight 4 invariant lagrangian of \eq{six}; at this order,
whose effects in the simulations is very tiny,  
we would therefore have to fit the independent parameter $b_4^{(1)}$, and this is at the moment beyond the reach
of our precision. 

Perturbatively, and up to the order discussed above, the correction to the partition function for a 
given world-sheet geometry due to the boundary terms is given by
\begin{equation}
\label{delZ}
\delta_{\mathrm{bound}}^{(1)} Z= 
- Z_{\mathrm{free}} \langle S_{b,2}^{(1)} \rangle~,
\end{equation}
where $Z_{\mathrm free}$ is the partition function obtained using the free action (the first term in
\eq{action})
\begin{equation}
\label{freeact}
S_{\mathrm{free}} = \frac{\sigma}{2} \int_{\Sigma} \left[(\partial_0 X)^2 + (\partial_1 X)^2\right] 
\end{equation}
and
\begin{equation}
\label{pertvev}
\begin{aligned}
\langle S_{b,2}^{(1)}\rangle &
= \frac{1}{Z^{\mathrm{free}}} 
\int DX\, S_{b,2}^{(1)}\, \ee^{-S_{\mathrm{free}}} 
=  \frac{1}{Z^{\mathrm{free}}} 
\int DX\, b_2  \int_{\partial \Sigma} (\partial_0\partial_1 X)^2 \, \ee^{-S_{\mathrm{free}}} 
\\
& = b_2\, \int_{\partial\Sigma} 
\lim_{\xi^\prime_0, \xi^\prime_1 \to \xi_0, \xi_1} \frac{\partial~}{\partial
\xi^\prime_0} \frac{\partial~}{\partial \xi^\prime_1}  
\frac{\partial~}{\partial \xi_0} \frac{\partial~}{\partial \xi_1} G(\vec
\xi^\prime,\vec \xi)~.
\end{aligned}
\end{equation}
In the last line we used Wick theorem to relate the result to the free
propagator $G(\vec \xi^\prime,\vec \xi)$
of the field $X$, which depends on the conditions imposed at the boundary $\partial\Sigma$.
In the following, we shall distinguish with a subscript $P$ or $W$ the boundary conditions corresponding to Polyakov loop correlators or Wilson loops.

In \cite{Aharony:2010cx} the cylinder geometry corresponding 
to the correlator of two Polyakov loops 
was considered, with the following result:
\beq
\label{polybound}
\langle S^{(1)}_{b,2} \rangle_P=-b_2\frac{\pi^3 L}{60 R^4} E_4(\ii\frac{L}{2 R})
\eeq
 where $R$ is the distance between
the two Polyakov loops and $L$ the length of the compactified time direction (i.e. the inverse temperature). 

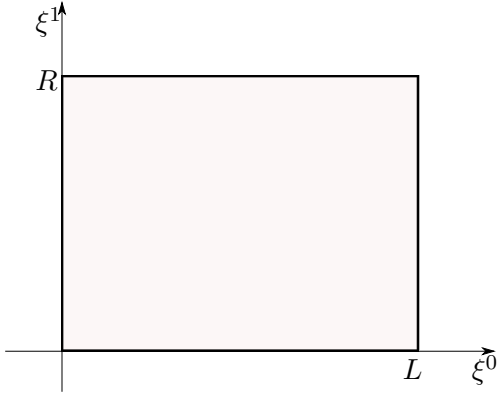
\begin{figure}
\begin{center}
\begingroup%
  \makeatletter%
  \providecommand\color[2][]{%
    \errmessage{(Inkscape) Color is used for the text in Inkscape, but the package 'color.sty' is not loaded}%
    \renewcommand\color[2][]{}%
  }%
  \providecommand\transparent[1]{%
    \errmessage{(Inkscape) Transparency is used (non-zero) for the text in Inkscape, but the package 'transparent.sty' is not loaded}%
    \renewcommand\transparent[1]{}%
  }%
  \providecommand\rotatebox[2]{#2}%
  \ifx\svgwidth\undefined%
    \setlength{\unitlength}{185bp}
    \ifx\svgscale\undefined%
      \relax%
    \else%
      \setlength{\unitlength}{\unitlength * \real{\svgscale}}%
    \fi%
  \else%
    \setlength{\unitlength}{\svgwidth}%
  \fi%
  \global\let\svgwidth\undefined%
  \global\let\svgscale\undefined%
  \makeatother%
  \begin{picture}(1,0.70885328)%
    \put(0,0){\includegraphics[width=\unitlength]{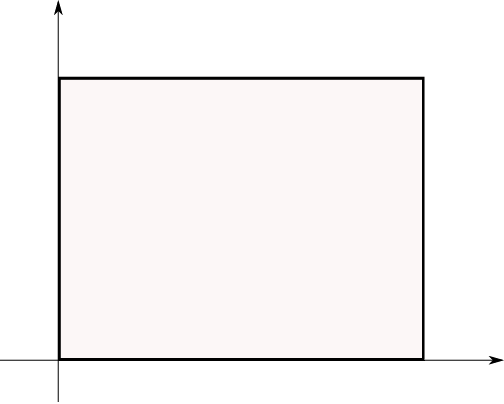}}%
    \put(0.061,0.74){\makebox(0,0)[lb]{\smash{$\xi^1$}}}%
    \put(0.81,0.03){\makebox(0,0)[lb]{\smash{$L$}}}%
    \put(0.061,0.62){\makebox(0,0)[lb]{\smash{$R$}}}%
    \put(0.95,0.03){\makebox(0,0)[lb]{\smash{$\xi^0$}}}%
  \end{picture}%
\endgroup%
\end{center}
\caption{The rectangular Wilson loop geometry that we consider.}
\label{fig:wl}
\end{figure}

Here we consider $\partial\Sigma$ to be a rectangular Wilson loop of sides $R$ and
$L$, as in figure \ref{fig:wl}. It has four components corresponding to the
sides of the rectangle. 

The Green function with Dirichlet boundary conditions on this loop is given by
\begin{equation}
G_W(\vec\xi^{'},\vec \xi)=\frac{4}{R L} \sum_{m,n=1}^{\infty} 
\frac{\sin{(\frac{n\pi \xi_1}{R})} \sin{(\frac{m \pi \xi_0}{L})} 
\sin{(\frac{n \pi \xi_1^{'}}{R})} \sin{(\frac{m \pi \xi_0^{'}}{L})}}{\pi^2 
\left( \frac{m^2}{L^2} + \frac{n^2}{R^2} \right)}
~.
\end{equation}

In order to compute eq. (\ref{pertvev}) we focus our attention on one side
of the Wilson loop, for example the one with coordinates $(0,\xi_1)$, with
$\xi_1\in [0,R]$. We have to compute the integral
\begin{equation}
\label{sideresult}
\begin{aligned}
\int_0^R d\xi^1 \partial_0^{'} \partial_1^{'} \partial_0 \partial_1
G_W(\vec\xi,\vec\xi^{'})|_{\xi_1^{'}=\xi_1 , \xi_0^{'}=\xi_0=0}
& =   \int_0^R d\xi^1 \, \frac{4 \pi^2}{R^3 L^3} \sum_{m,n=1}^{\infty}
\frac{m^2 n^2 (\cos{(\frac{n \pi \xi_1}{R})})^2  }{ ( \frac{m^2}{L^2} +
\frac{n^2}{R^2} )}\\
&  =  \frac{2 \pi^2}{R^2 L^3} \sum_{m,n=1}^{\infty}
\frac{m^2 n^2}{\left( \frac{m^2}{L^2} + \frac{n^2}{R^2} \right)}~. 
\end{aligned}
\end{equation}

With some manipulations we can rewrite the sum in eq. (\ref{sideresult}) as
follows:
\begin{equation}
\sum_{m,n=1}^{\infty} \frac{m^2 n^2}{( \frac{m^2}{L^2} + \frac{n^2}{R^2} )}
= 
R^2 \sum_{m=1}^{\infty} m^2 \sum_{n=1}^{\infty} 
\left(1-\frac{m^2}{L^2} \frac{1}
{\left(
\frac{m^2}{L^2} + \frac{n^2}{R^2} \right)}\right)\\
= 
-\frac{\pi R^3 }{2L} \sum_{m=1}^{\infty} m^3
\coth{\left(\frac{m \pi R}{L}\right)}~,
\end{equation}
where we used eq. (\ref{finitesum}) and regularized
the divergent sum using \eq{zeta}, taking into account that $\zeta(0) = -1/2$. 

Introducing $q=\exp (-2 \pi R/L)$ we can rewrite the above expression as
\begin{equation}
\begin{aligned}
\sum_{m,n=1}^{\infty} \frac{m^2 n^2}{( \frac{m^2}{L^2} + \frac{n^2}{R^2} )}
& = -\frac{\pi R^3 }{2L} \sum_{m=1}^{\infty} m^3 \left(1+\frac{2 q^m}{1 - q^m}
\right) = 
-\frac{\pi R^3 }{2L}\left\{\zeta(-3) +2 \sum_{m=1}^{\infty} \frac{m^3
q^m}{1-q^m})\right\}\\
& = -\frac{\pi R^3 }{2L}\zeta(-3) E_4\left(\ii\frac{R}{L}\right)~.
\end{aligned}
\end{equation}
Here we used \eq{zeta} to regularize the divergent series and \eq{sume4bis} to
rewrite the remaining sum in terms of the Eisenstein series $E_4$. 

Substituting the last expression in eq. (\ref{sideresult}), and using the explicit
value $\zeta(-3) = 1/120$, we find thus that the 
contribution to the boundary term \eq{pertvev} from the $(0,\xi_1)$ side of
the rectangle is given by
\begin{equation}
\label{onesideresult}
\int_0^R d\xi^1 \partial_0^{'} \partial_1^{'} \partial_0 \partial_1
G_W(\vec\xi,\vec\xi^{'})|_{\xi_1^{'}=\xi_1 , \xi_0^{'}=\xi_0=0} =-\frac{\pi^3
R}{120 L^4} E_4\left(\ii\frac{R}{L}\right)~.
\end{equation}
The simmetry of the boundary implies that the contribution from
the other side in the same direction, the one parametrized by $(L,\xi_1)$,
with $\xi_1\in [0,R]$, must be the same. The contribution from the two sides of
the rectangle in the orthogonal direction are obtained exchanging $R$ and $L$.
Altogether we have thus
\begin{equation}
\label{wilsbound}
\langle S^{(1)}_{b,2} \rangle_W=-b_2 \frac{\pi^3}{60}\left[\frac{R}{L^4}
E_4\left(\ii\frac{R}{L}\right) + \frac{L}{R^4} E_4\left(\ii\frac{L}{R}\right)\right]~.
\end{equation}

\section{Numerical results}
In order to test our prediction for the boundary terms we performed a set of simulations in  
the three-dimensional $\Z_2 $ gauge model. We used the following strategy. We first 
evaluated a set of Polyakov loop correlators in the low temperature regime. 
Comparing our results with eq. (\ref{polybound}) we could verify the correctness of 
the effective string prediction and extract an estimate of the coefficient $b_2$. 
To check the correct scaling behaviour of $b_2$ we performed this analysis for three 
different values of the bare coupling $\beta$.
Then we inserted this estimate for $b_2$ into eq. (\ref{wilsbound}) to predict the boundary correction in the Wilson loop case and compared it with the simulation results. In this last comparison there was no free parameter to fit.
The details of the simulation algorithm that we employed can be found in \cite{{Caselle:2002ah},{Billo:2011fd}}. 
As a basic update mechanism we used a multispin coded version of the standard metropolis algorithm.
Let us now describe the various steps of our analysis.

\vskip 0.2cm
\noindent
{\sl Polyakov loop correlators}

In order to eliminate the non-universal perimeter and constant terms from the 
expectation value of Polyakov loop correlators $P(R,L)$ (where $L$ is the length of the two loops and
$R$ their distance) we measured the following ratio: 
\beq
\Rp(R,L)=\frac{P(R+1,L)}{P(R,L)}~.
\eeq
Due to the peculiar nature of our algorithm, based on the dual transformation to the 3d spin Ising model, this ratio can be evaluated for large values of $R$ and $L$ with very high precision.

The effective string prediction for this observable reads, up to the second loop order,
\beq
\Rp(R,L)=\ee^{-\sigma L}\, \frac{\eta(\ii\frac{L}{2 R})}{\eta(\ii\frac{L}{2R+2})} 
\left(1+F_2(R+1,L)+F_P(R+1,L)-F_2(R,L)-F_P(R,L)\right)~,
\eeq
where $\eta$ is Dedekind's function, $\sigma$ is the string tension, $F_2$  the standard two loop effective string contribution to the Polyakov loop correlator (see for instance \cite{Caselle:2005xy})
\beq
F_2(R,L)= \frac{ \pi^2 L}{1152 R^3 \sigma} (2 E_4(\ii\frac{L}{2 R})-E_2^2(\ii\frac{L}{2R}))
\eeq
and $F_P=\langle S^{(1)}_{b,2} \rangle_P$ is the leading correction coming from the boundary \cite{Aharony:2010cx}
given in \eq{polybound},
which we report here for completeness:
\beq
F_P(R,L)=-b_2\frac{\pi^3 L}{60 R^4} E_4(\ii\frac{L}{2 R})~.
\label{Fpagain}
\eeq
We can use the fact that $\sigma $ is known with very high precision to define a new observable 
which allows to isolate the boundary term:
\beq
\Rp^\prime(R,L)=\ee^{\sigma L} \frac{\eta(\ii\frac{L}{2R+2})}{\eta(\ii\frac{L}{2 R})} \Rp(R,L)-F_2(R+1,L)+F_2(R,L) - 1
\eeq
{}From \eq{Fpagain} we expect for this observable, when $L\gg2R$,
\beq
\Rp^\prime(R,L)=b_2 \frac{L \pi^3}{15 R^5} \label{rp}
\eeq
We measured this observable for three values of the bare coupling $ \beta $.
Table \ref{simulation} displays the values of $ \beta $ and the corresponding string tension,
as well as the values of $ L $ and of the inverse of the critical temperature $1/{T_c}$, for the three data sets.  We always chose $L\gg 1/T_c$ 
to avoid the additional complication of dealing with finite temperature correction terms. 

\begin{table}[ht]
\centering
\begin{tabular}{|ccccc|}
\hline
data set & $ \beta $ & $ L $ & $ \sigma $ & $1/T_c$   \\
\hline
 1  & 0.743543 & 68  &  0.0228068(15) & 5 \\
 2  & 0.751805 & 100 &  0.0105255(11) & 8 \\
 3  & 0.754700 & 125 &  0.0067269(17) & 10 \\
\hline
\end{tabular}
\caption{Some informations on the data sample}
\label{simulation}
\end{table}

We report in table \ref{polres}  the numerical results for the various datasets.
\begin{table}[ht]
\centering
\begin{tabular}{|cc|cc|cc|}
\hline
R & data set 1 &  R & data set 2 & R &  data set 3  \\
\hline
 11  & 0.00611(15) & 16 & 0.00515(18) & 20 & 0.00446(13) \\
 12  & 0.00431(15) & 17 & 0.00395(18) & 21 & 0.00355(13) \\
 13  & 0.00290(16) & 18 & 0.00334(18) & 22 & 0.00304(13) \\
 14  & 0.00182(16) & 19 & 0.00207(18) & 23 & 0.00226(14) \\
 15  & 0.00159(16) & 20 & 0.00132(18) & 24 & 0.00190(14)  \\
 16  & 0.00092(16) & 21 & 0.00119(18) & 25 & 0.00162(14) \\
 17  & 0.00088(16) & 22 & 0.00133(18) & 26 & 0.00127(14) \\
 18  & 0.00062(16) & 23 & 0.00083(18) & 27 & 0.00134(14) \\
 19  & 0.00020(16) & 24 & 0.00071(18) & 28 & 0.00104(14) \\
 20  & 0.00016(16) & 25 & 0.00088(18) & 29 & 0.00074(14) \\
     &             & 26 & 0.00067(18) & 30 & 0.00060(14) \\
     &             &    &             & 31 & 0.00078(14) \\
     &             &    &             & 32 & 0.00062(14) \\
     &             &    &             & 33 & 0.00053(14) \\
     &             &    &             & 34 & 0.00033(14) \\
\hline
\end{tabular}
\caption{Numerical results for $ \Rp^\prime(R,L) $}
\label{polres}
\end{table}
We fitted the numerical data in table \ref{polres} with the function in eq. (\ref{rp}), where the only free parameter is $ b_2 $. The results of the fit are given in table \ref{bres}. 
For each value of $\beta$ we chose the lowest value for $ R $ in the fit to be roughly twice
the value of the inverse critical temperature. Previous analysis on this model \cite{Caselle:2005xy} showed that
this threshold is well within the regime of validity of the effective string picture.
Looking at table \ref{bres} we see that in the fits obtained from all the data sets the $\chi^2$ values are of order one. This strongly supports the correctness of eq. (\ref{polybound}), obtained in \cite{Aharony:2010cx}. It is interesting to note that the value of $b_2$ changes dramatically with $\beta$. This is due to the fact that $b_2$ has the dimensions of  $(\mathrm{length})^3$. In fact, a remarkable consistency check of our results is that the adimensional combination $b_2 \sqrt{\sigma}^3 $ is almost constant. This combination can be used to define the continuum limit value of $b_2$ which can thus be considered as
 a new physical scale of the model, on the same ground as the string tension $\sigma$  or the glueball mass $m_g$. A slight scaling deviation is still visible between the  data set 2 and the data set 3. Since we have no hint of the behaviour of these scaling corrections we propose as our final estimate for the continuum limit value of the combination $ b_2 \sqrt{\sigma}^3$
the value obtained in the data set 3 (which is the nearest to the continuum limit) with a systematic error equal to the difference (within statistical errors)
between the value obtained in the data set 3 and that of data set 2. In this way we obtain  $ b_2 \sqrt{\sigma}^3 \sim 0.032(2)$. 
 It is interesting to notice
 that our result for $ b_2 $ in the $ 3d $ $ \Z_2 $ gauge model is very similar (but ten times more precise) to the same parameter measured in the $ 3d $ $ SU(2) $ 
 gauge model in \cite{Brandt:2010bw}.

\begin{table}[ht]
\centering
\begin{tabular}{|cccc|}
\hline
data set & $ b_2 $ & $ b_2 \sqrt{\sigma}^3 $  & $ \chi^2 $    \\
\hline
 1  & 7.25(15) & 0.0250(5) & 1.2 \\
 2  & 26.8(8)  & 0.0289(9) & 1.8 \\
 3  & 57.9(12) & 0.0319(7) & 1.3 \\
\hline
\end{tabular}
\caption{Some informations on the data sample}
\label{bres}
\end{table}

\vskip 0.2cm
\noindent
{\sl Wilson loops.}

Using the above value of $b_2$ we can now test  the leading boundary contribution 
to the Wilson loop reported  in eq. (\ref{wilsbound}). Since  we know $ \sigma $ and $ b_2 $ from independent simulations, 
there are no free parameters left, thus we may perform, so to speak, an "absolute" test of our predictions.

In order to eliminate the non-universal perimeter and costant terms we introduce (as in \cite{Billo:2011fd}) the following ratio of Wilson loops : 
\beq
\label{Rw}
\Rw(L,R)=\frac{W(L,R)}{W(L+1,R-1)} 
\eeq
where $W(R,L)$ denotes the expectation value of a Wilson loop of size $R\times L$
(see fig. \ref{fig:wl}).
Then in order to eliminate the dominant confining term, using again the fact the $\sigma$ is known with very high precision 
we define
\beq
\Rw^{'}(L,L u)=\Rw(L,u L)-\exp \{-\sigma (1+L(1-u)) \}~,
\eeq
where we introduced the asymmetry parameter $ u= R/L$ (with $R>L $).

In table \ref{wilsres} we report the numerical results for this observable
evaluated for the same three values of $ \beta $ reported in table \ref{simulation}.

\begin{table}[ht]
\centering
\begin{tabular}{|cc|cc|cc|}
\hline
R & data set 1 &  R & data set 2 & R &  data set 3  \\
\hline
 12  & -0.00642(8) & 12 & -0.00687(4) & 21 & -0.00356(10) \\
 15  & -0.00468(9) & 15 & -0.00512(9) & 24 & -0.00292(10) \\
 18  & -0.00376(10) & 18 & -0.00414(10) & 27 & -0.00246(11) \\
 21  & -0.00285(11) & 21 & -0.00317(10) & 30 & -0.00223(12) \\
 24  & -0.00247(11) & 24 & -0.00282(11) & 33 & -0.00182(12)  \\
 27  & -0.00204(12) &    &              &    &  \\

\hline
\end{tabular}
\caption{Numerical results for $\Rw^{'}(L,\frac{4}{3} L) $}
\label{wilsres}
\end{table}

\begin{figure}[htm]
\centering{\includegraphics[width=10cm]{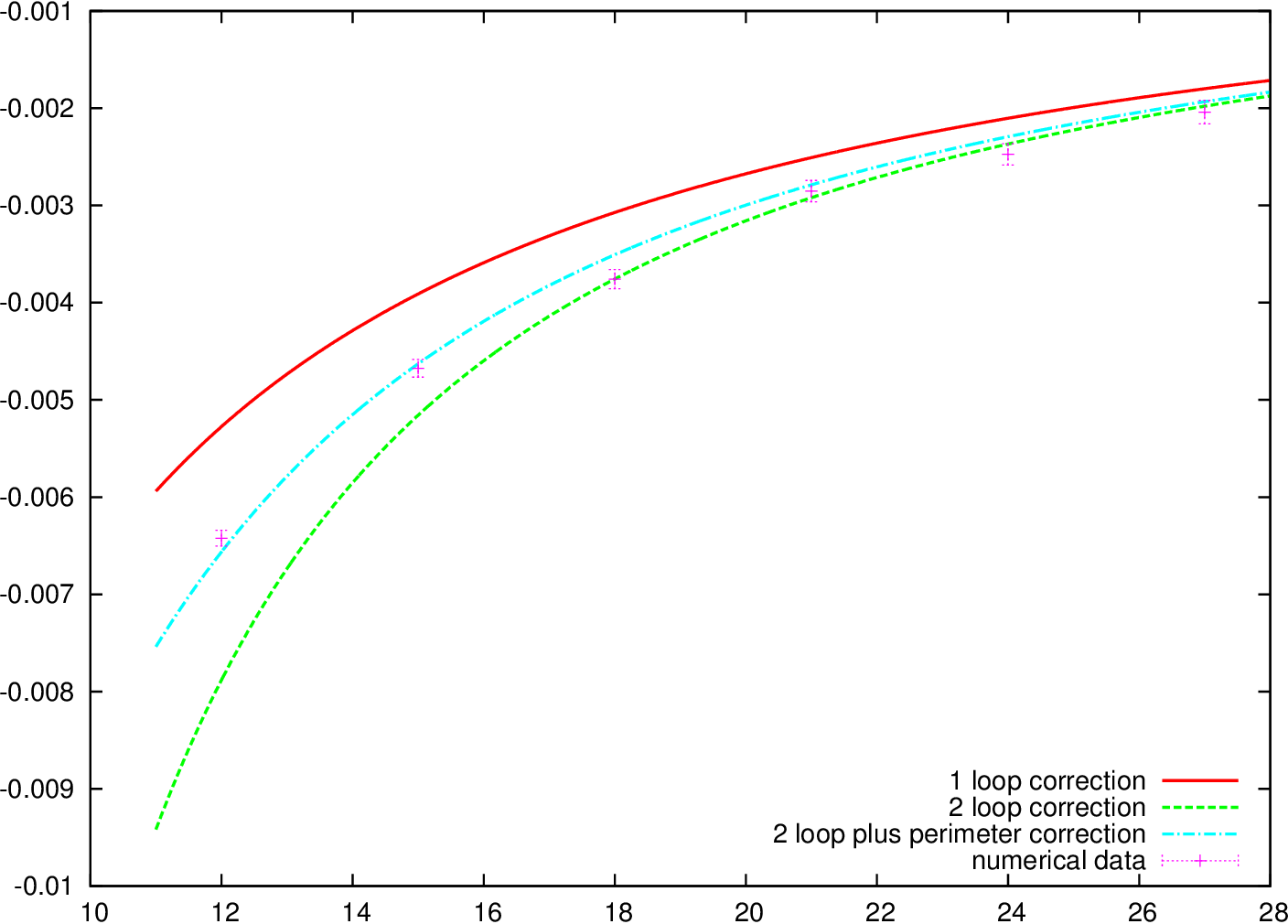}}
\caption{$ \Rw^{'}(L,L \frac{4}{3}) $ at $\beta=0.743543$}
\label{beta74}
\end{figure}

\begin{figure}[htm]
\centering{\includegraphics[width=10cm]{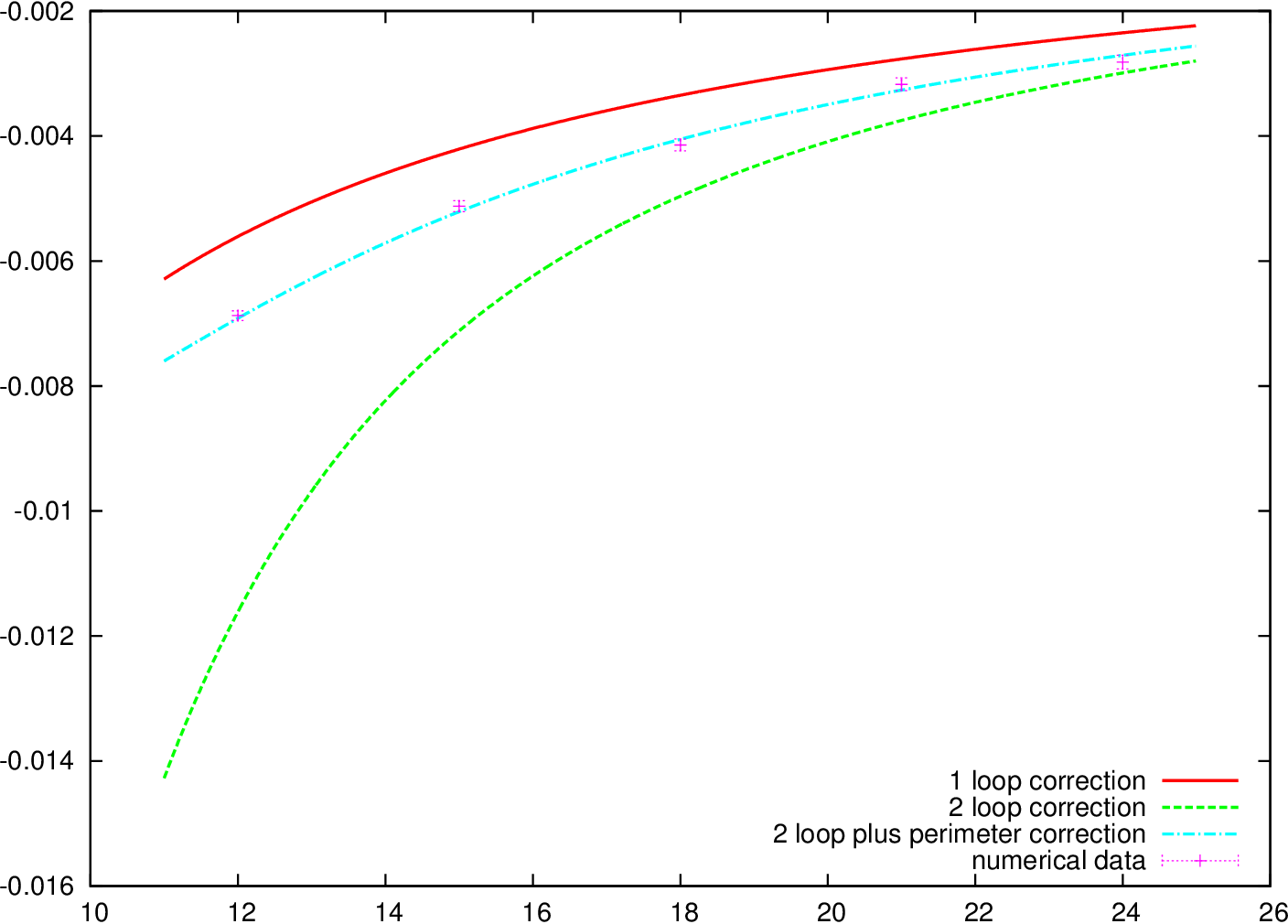}}
\caption{$ \Rw^{'}(L,L \frac{4}{3}) $ at $\beta=0.751805 $ }
\label{beta751}
\end{figure}

\begin{figure}[htm]
\centering{\includegraphics[width=10cm]{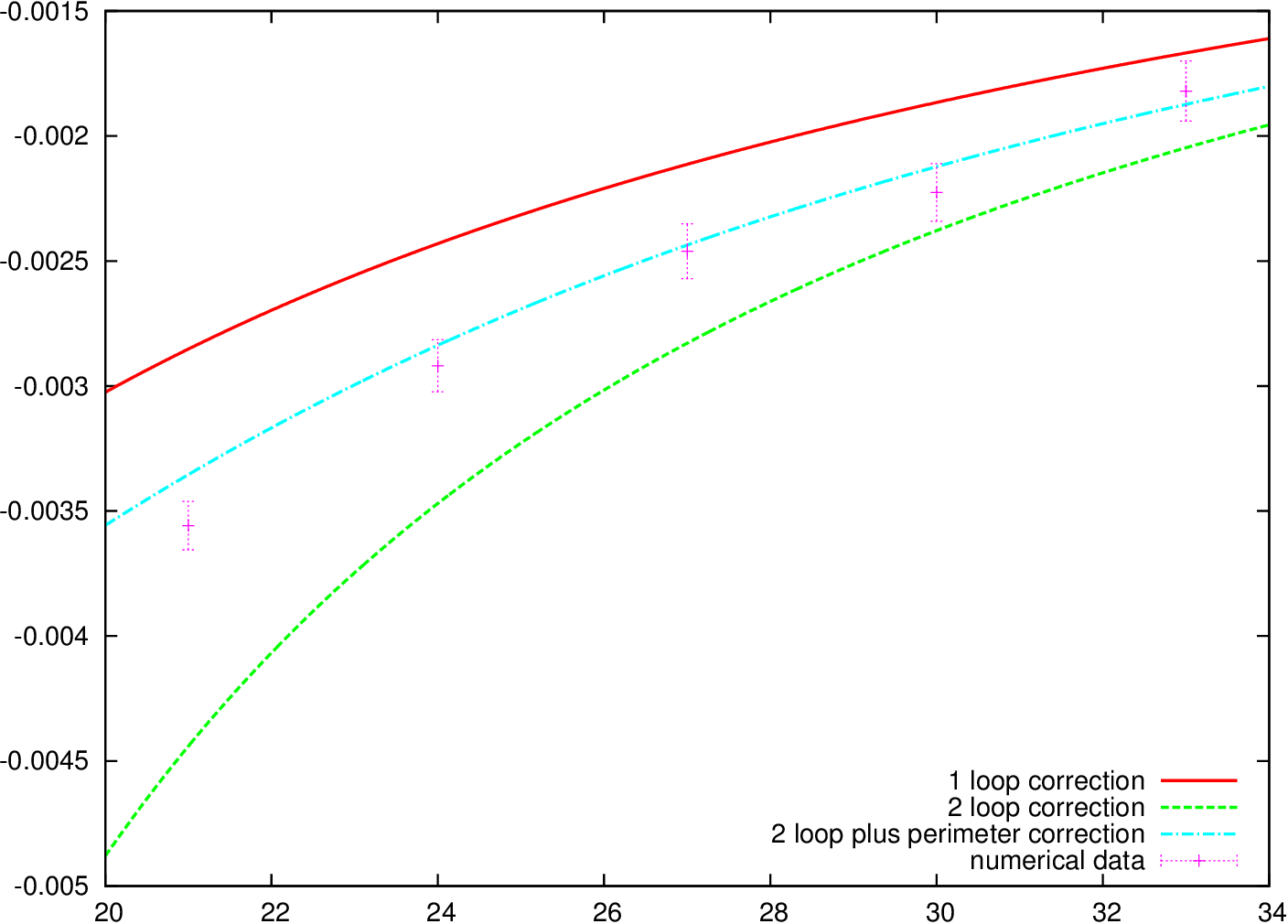}}
\caption{$ \Rw^{'}(L,\frac{4}{3} L) $ at $\beta=0.754700 $}
\label{beta754}
\end{figure}

In fig.s \ref{beta74},\ref{beta751} and \ref{beta754} we 
plotted the effective string prediction including the one loop and two loop corrections from the bulk 
string action,  which we report here for completeness (see \cite{Billo:2011fd} for a detailed derivation) 
\begin{equation}
 \label{wlngexp}
W(L,R) = \ee^{-\sigma RL+p(R+L)+k}\, \left(\frac{\eta(\ii
u)}{\sqrt{L}}\right)^{-\frac{d-2}{2}}\left\{1 + \frac{\cL_2(u)}{\sigma RL} + \ldots\right\}~,
\end{equation}
with
\begin{equation}
\cL_2(u)
=\left(\frac{\pi}{24}\right)^2 \left[2(d-2)\, u^2 E_4\left(\ii u\right)- 
\frac{(d-2)(d-6)}{2} E_2\left(\ii u\right)E_2\left(\ii/u\right) \right]~.
\label{2lng}
\end{equation}
We also report in the figures the curve obtained adding to the bulk string terms  
the leading correction coming from the boundary, given in eq. (\ref{wilsbound}), and
the numerical results of our simulations.


The agreement between the data and the prediction is impressive and, with the precision of our data, the boundary correction is absolutely necessary to correctly describe them. 
Indeed, looking at the similar analysis discussed in \cite{Billo:2011fd} (which did not keep into account boundary terms) it is clear 
that the boundary correction discussed in this paper exactly fills the gap between data and prediction 
found, for instance, in fig. 2 of \cite{Billo:2011fd}.

\section{Conclusions}
\label{sec:conclusions}
In this paper we used the requirement of Lorentz invariance to constrain
the effective boundary action of the confining string which describes 
the  behaviour of the Wilson loops and Polyakov
correlators in any confining gauge theory. We wrote explicitly all the possible
Lorentz invariants up to terms of scaling 8. We also presented the
analytic calculation of the contribution of the leading term of 
the boundary action to the mentioned observables (the calculation for the Polyakov correlator was already done in  \cite{Aharony:2010cx}). Although the boundary term depends only on (derivatives) of string coordinates evaluated on the boundary, its  contribution due to quantum fluctuations depends on the whole shape of the observable and can be expressed in a closed form in terms of the Eisenstein series $E_4$.
The coefficient $b_2$ in front of this boundary term has  the dimensions of 
$({\rm length)^3}$ , thus it defines a new physical scale in any confining gauge theory. It is interesting to observe that in the derivative expansion of the string action, $b_2$ is the first term where one can find a dependence on the gauge group, while all the terms of lower scaling are completely fixed and are the same for any 3D gauge theory. 
We evaluated the coefficient $b_2$ in the three-dimensional $\Z_2$ gauge 
model  using the data of  Monte Carlo simulations for the Polyakov correlator 
corresponding to three different values of the bare coupling in order to check the correct scaling behaviour of $b_2$. We then inserted this estimate to predict the boundary correction in the Wilson loop case and compared it with the data of the numerical simulation. Although there are no adjustable parameters, the agreement 
between the prediction and  the data is impressive, as it is clearly visible in fig.s \ref{beta74}, \ref{beta751} and \ref{beta754}.
\vskip 1cm
\noindent {\large {\bf Acknowledgements}}
\vskip 0.2cm
The MC simulations were performed at the INFN Pisa GRID DATA center and on
the INFN cluster CSN4. We thank Mattia Bruno for discussions, and in particular for pointing 
out a few misprints in the first version of this paper.

\vskip 1cm
\appendix

\section{Useful formulae}
\label{app:formulae}
In section \ref{sec:bcwilson} we make use of the following series for the
hyperbolic cotangent:
\begin{equation}
\label{finitesum}
\sum_{n=1}^{\infty} \frac{1}{ ( \frac{m^2}{L^2} + \frac{n^2}{R^2} )}=\frac{\pi R
L}{2m} \coth{(\frac{m \pi R}{L})}-\frac{L^2}{2 m^2}~.
\end{equation}
We also encounter divergent sums which we regularize by means of the zeta
function
\begin{equation}
\label{zeta}
\zeta(s) \equiv \sum_{n=1}^{\infty} n^{-s}~.
\end{equation}
Dedekind's $\eta$-function is defined as
\begin{equation}
\label{defeta}
\eta(\tau) = q^{\frac{1}{24}} \prod_{n=1}^\infty (1 - q^n)~,
\end{equation}
where $q=\exp(2\pi\ii\tau)$.
The Eisenstein series $E_{2k}$, with $k=1,2,\ldots$, are defined by
\begin{equation}
\label{defeis} 
E_{2k}(\tau) = 1 + \frac{2}{\zeta(1 - 2k)} \sum_{n=1}^\infty \sigma_{2k-1}(n)
q^{n}~,
\end{equation}
where $\sigma_p(n)$ is the sum of the $p$-th powers of the divisors of $n$:
\begin{equation}
\label{defsigma}
\sigma_p(n) = \sum_{m|n} m^p~. 
\end{equation}

In section \ref{sec:bcwilson} we encounter the sum 
\begin{equation}
 \label{sume4}
2 \sum_{m=1}^\infty \frac{m^3 q^m}{1 - q^m} = 2 \sum_{m=1}^\infty
\sum_{s=1}^\infty m^3 q^{m s} = 2 \sum_{n=1}^\infty \sigma_3(n) q^n~, 
\end{equation}
where in the last step we changed summation variables to $n=m s$ and $m$, so
that $m$ is a divisor of $n$, and employed the definition \eq{defsigma}. From
\eq{defeis} we have then
\begin{equation}
 \label{sume4bis}
 2 \sum_{m=1}^\infty \frac{m^3 q^m}{1 - q^m} = \zeta(-3) \left(E_4(\tau)
- 1\right)~.
\end{equation}
In section \ref{sec:bcwilson} the parameter $\tau$ is in fact $\ii R/L$. 

\section{Boundary Lorentz invariants of higher scaling}
\label{appb}
In this appendix we list the Lorentz-invariant  terms 
of scaling 4 and 6 which contribute to the boundary action
of the open string. Since the infinitesimal transformation
(\ref{rota}) when applied to a generic term $\partial^mX^n$ 
keeps invariant the scaling $m-n$, it suffices to list 
the set of all possible terms of definite scaling with arbitrary 
coefficients and then solve the recursion relations 
dictated by the request of invariance. For instance in $d=3$ 
there are three different sequences of terms of scaling 4, namely
\beq
\label{seq4}
\begin{aligned}
\cL_{b,4}[d=3]=\sum_n\Big[\alpha_n(\partial_1X)^{2n}\left(\partial_0^2\partial_1X\right)^2+&
\beta_n(\partial_1X)^{2n+1}\partial_0^2\partial_1X
\left(\partial_0\partial_1X\right)^2\\+&
\gamma_n(\partial_1X)^{2n+2}\left(\partial_0\partial_1X\right)^4\Big]\,.
\end{aligned}
\eeq
Demanding $\delta\cL_{b,4}=0$ generates the following set of recursion
relations
\beqa
0=&(n+2)\alpha_n+(n+1)\alpha_{n+1}\,,\\
0=&4\alpha_{n+1}+(2n+7)\beta_n+(2n+3)\beta_{n+1}\,,\\
0=&\beta_{n+1}+(n+5)\gamma_n+(n+2)\gamma_{n+1}\,.
\eeqa 
They can be solved in terms of a single free parameter $c=b^4_1$, yielding
\beq
\alpha_n=(-1)^n(n+1)c\,,\beta_n=-2(-1)^n(n+1)(n+2)c\,,
\gamma_n=\frac23(-1)^n(n+1)(n+2)(n+3)c\,.
\eeq
Inserting these solutions in (\ref{seq4}) gives at once (\ref{six}).
It is not difficult to prove that its generalization to $d$  
space-time dimensions is
\beq
\begin{aligned}
\cL_{b,4}=&\frac{\partial_0^2\partial_1X\cdot\partial_0^2\partial_1X}
{1+\partial_1X\cdot \partial_1X }-
\frac{\left(\partial_0^2\partial_1X\cdot\partial_1X\right)^2+
4\left(\partial_0^2\partial_1X\cdot\partial_0\partial_1X\right)
\left(\partial_0\partial_1X\cdot\partial_1X\right)}
{\left(1+\partial_1X\cdot \partial_1X\right)^2 }\\
+&4\frac{\left(\partial_0\partial_1X\cdot\partial_1X\right)^2
   \left[\partial_1X\cdot\partial_0^2\partial_1X +
\partial_0\partial_1X\cdot\partial_0\partial_1X \right]}
{\left(1+\partial_1X\cdot \partial_1X\right)^3}
-4\frac{\left(\partial_1X\cdot\partial_0\partial_1X\right)^4}
{\left(1+\partial_1X\cdot\partial_1X\right)^4}~.
\end{aligned}
\eeq

Using the same procedure it is possible to show that there are in total five 
Lorentz invariants of scaling 6;  three of them 
 cannot be written as products of 
invariants of lower scaling. We list them for the sake of completeness:

\beq
\begin{aligned}
\cL_{b,6}^1=& 
 \frac{1}{\left(1+\partial_1X\cdot\partial_1X\right)^3}\Bigl\{
 \partial_0^3\partial_1X\cdot\partial_0\partial_1X
\left[\partial_0\partial_1X\cdot\partial_0\partial_1X
\left(1+\partial_1X\cdot\partial_1X\right)
-\left(\partial_0\partial_1X\cdot\partial_1X\right)^2\right]
\\
- & \left(\partial_0^3\partial_1X\cdot\partial_1X\right)
\left(\partial_0\partial_1X\cdot\partial_0\partial_1X\right)
\left(\partial_0\partial_1X\cdot\partial_1X
\right)}{\left(1+\partial_1X\cdot\partial_1X\right)^3\Bigr\}     \\
+&\frac{\left(\partial_0^3\partial_1X\cdot\partial_1X\right)
\left(\partial_0\partial_1X\cdot\partial_1X
\right)^3}{\left(1+\partial_1X\cdot\partial_1X\right)^4}     
-3 \frac{\left(\partial_0^2\partial_1X\cdot\partial_0\partial_1X\right)
\left(\partial_0\partial_1X\cdot\partial_0\partial_1X\right)
\left(\partial_0\partial_1X\cdot\partial_1X
\right)}{\left(1+\partial_1X\cdot\partial_1X\right)^3}      \\
-3 & \frac{\left(\partial_0^2\partial_1X\cdot\partial_1X\right)
\left(\partial_0\partial_1X\cdot\partial_0\partial_1X\right)^2}
{\left(1+\partial_1X\cdot\partial_1X\right)^3}     
+3 \frac{\left(\partial_0^2\partial_1X\cdot\partial_0\partial_1X\right)
\left(\partial_0\partial_1X\cdot\partial_1X\right)^3
}{\left(1+\partial_1X\cdot\partial_1X\right)^4}  \\ 
+6 &\frac{
\left(\partial_0\partial_1X\cdot\partial_1X
\right)^6}{\left(1+\partial_1X\cdot\partial_1X\right)^6}    
+9 \frac{\left(\partial_0\partial_1X\cdot\partial_0\partial_1X\right)
\left(\partial_0^2\partial_1X\cdot\partial_1X\right)
\left(\partial_0\partial_1X\cdot\partial_1X
\right)^2}{\left(1+\partial_1X\cdot\partial_1X\right)^4}\\
-6 &       \frac{\left(\partial_0^2\partial_1X\cdot\partial_1X\right)
\left(\partial_0\partial_1X\cdot\partial_1X
\right)^4}{\left(1+\partial_1X\cdot\partial_1X\right)^5}
+6 \frac{\left(\partial_0\partial_1X\cdot\partial_0\partial_1X\right)^2
\left(\partial_0\partial_1X\cdot\partial_1X
\right)^2}{\left(1+\partial_1X\cdot\partial_1X\right)^4} \\     
-12 &\frac{\left(\partial_0\partial_1X\cdot\partial_0\partial_1X\right)
\left(\partial_0\partial_1X\cdot\partial_1X
\right)^4}{\left(1+\partial_1X\cdot\partial_1X\right)^5}\,,
\end{aligned}
\eeq 

\beq
\begin{aligned}
\cL_{b,6}^2=& \frac{\partial_0^3\partial_1X\cdot\partial_0^3\partial_1X}
{1+\partial_1X\cdot\partial_1X}-\frac{\left(\partial_0^3\partial_1X\cdot
\partial_1X\right)^2}{\left(1+\partial_1X\cdot\partial_1X\right)^2}
-6\frac{\left(\partial_0^3\partial_1X\cdot
\partial_0^2\partial_1X\right)\left(\partial_1X\cdot\partial_0\partial_1X
\right)}{\left(1+\partial_1X\cdot\partial_1X\right)^2}\\
-6&\frac{\left(\partial_0^3\partial_1X\cdot
\partial_0\partial_1X\right)\left(\partial_1X\cdot\partial_0^2\partial_1X
\right)}{\left(1+\partial_1X\cdot\partial_1X\right)^2}
+12  \frac{\left(\partial_0^3\partial_1X\cdot\partial_1X\right)
\left(\partial_0^2\partial_1X\cdot\partial_1X\right)
\left(\partial_0\partial_1X\cdot\partial_1X
\right)}{\left(1+\partial_1X\cdot\partial_1X\right)^3}      \\
+12&  \frac{\left(\partial_0^3\partial_1X\cdot\partial_0\partial_1X\right)
\left(\partial_0\partial_1X\cdot\partial_1X\right)^2
}{\left(1+\partial_1X\cdot\partial_1X\right)^3}
 -12  \frac{\left(\partial_0^3\partial_1X\cdot\partial_1X\right)
\left(\partial_0\partial_1X\cdot\partial_1X\right)^3
}{\left(1+\partial_1X\cdot\partial_1X\right)^4}\\
+9&\frac{\left(\partial_0^2\partial_1X\cdot\partial_1X\right)^2
\left(\partial_0\partial_1X\cdot\partial_0\partial_1X\right)
}{\left(1+\partial_1X\cdot\partial_1X\right)^3}
+9 \frac{\left(\partial_0^2\partial_1X\cdot\partial_0^2\partial_1X\right)
\left(\partial_0\partial_1X\cdot\partial_1X\right)^2
}{\left(1+\partial_1X\cdot\partial_1X\right)^3}\\
+18& \frac{\left(\partial_0^2\partial_1X\cdot\partial_0\partial_1X\right)
\left(\partial_0^2\partial_1X\cdot\partial_1X\right)
\left(\partial_0\partial_1X\cdot\partial_1X
\right)}{\left(1+\partial_1X\cdot\partial_1X\right)^3}
-36  \frac{\left(\partial_0^2\partial_1X\cdot\partial_1X\right)^2
\left(\partial_0\partial_1X\cdot\partial_1X
\right)^2}{\left(1+\partial_1X\cdot\partial_1X\right)^4}\\
-36&  \frac{\left(\partial_0\partial_1X\cdot\partial_1X
\right)^2\left[\left(\partial_0^2\partial_1X\cdot\partial_0\partial_1X\right)
\left(\partial_0\partial_1X\cdot\partial_1X
\right)+\left(\partial_0\partial_1X\cdot\partial_0\partial_1X
\right)\left(\partial_0^2\partial_1X\cdot\partial_1X
\right)\right] }{\left(1+\partial_1X\cdot\partial_1X\right)^4}\\
+36&\frac{\left(\partial_0\partial_1X\cdot\
\partial_1X\right)^4\left[
2\partial_0^2\partial_1X\cdot\partial_1X+
\partial_0\partial_1X\cdot\partial_0\partial_1X\right]
   }{\left(1+\partial_1X\cdot\partial_1X\right)^5}
-36\frac{\left(\partial_0\partial_1X\cdot\partial_1X\right)^6}{\left(1+\partial_1X\cdot\partial_1X\right)^6}\,,
\end{aligned}
\eeq 

\beq
\begin{aligned}
\cL_{b,6}^3=& \frac{\left(\partial_0^2\partial_1X\cdot\partial_0\partial_1X\right)^2}{\left(1+\partial_1X\cdot\partial_1X\right)^2}+ 4\frac{\left(\partial_0\partial_1X\cdot\partial_1X\right)^6}{\left(1+\partial_1X\cdot\partial_1X\right)^6} \\
- 4 &\frac{\left(\partial_0\partial_1X\cdot\
\partial_1X\right)^4\left[
\partial_0^2\partial_1X\cdot\partial_1X+
2\partial_0\partial_1X\cdot\partial_0\partial_1X\right]
   }{\left(1+\partial_1X\cdot\partial_1X\right)^5}\\
+4&\frac{\left(\partial_0\partial_1X\cdot\partial_1X
\right)^2\left[\left(\partial_0^2\partial_1X\cdot\partial_0\partial_1X\right)
\left(\partial_0\partial_1X\cdot\partial_1X
\right)+\left(\partial_0\partial_1X\cdot\partial_0\partial_1X
\right)\left(\partial_0^2\partial_1X\cdot\partial_1X
\right)\right] }{\left(1+\partial_1X\cdot\partial_1X\right)^4}\\
+4& \frac{\left(\partial_0\partial_1X\cdot\partial_0\partial_1X\right)^2
\left(\partial_0\partial_1X\cdot\partial_1X
\right)^2}{\left(1+\partial_1X\cdot\partial_1X\right)^4}
-4\frac{\left(\partial_0^2\partial_1X\cdot
\partial_0\partial_1X\right) \left(\partial_0\partial_1X\cdot
\partial_0\partial_1X\right)\left(\partial_0\partial_1X\cdot
\partial_1X\right)  }{\left(1+\partial_1X\cdot\partial_1X\right)^3}\\
+&\frac{\left(\partial_0^2\partial_1X\cdot
\partial_1X\right)^2 \left(\partial_0\partial_1X\cdot
\partial_1X\right)^2 }{\left(1+\partial_1X\cdot\partial_1X\right)^4}
-2\frac{\left(\partial_0^2\partial_1X\cdot
\partial_0\partial_1X\right) \left(\partial_0^2\partial_1X\cdot
\partial_1X\right)\left(\partial_0\partial_1X\cdot
\partial_1X\right)  }{\left(1+\partial_1X\cdot\partial_1X\right)^3}\,.
\end{aligned}
\eeq 

\providecommand{\href}[2]{#2}\begingroup\raggedright\endgroup

\end{document}